\documentclass[preprint,prd,noshowpacs,nofootinbib]{revtex4}
\usepackage{amssymb}
\usepackage{amsmath}
\usepackage{xcolor}
\usepackage{graphicx}  

\begin{document}

\title{Non-Abelian Firewall 
\footnote{Essay written for the Gravity Research Foundation 2020 Awards for Essays on Gravitation. \\}}

\author{Douglas Singleton} 
\email{dougs@mail.fresnostate.edu}
\affiliation{Physics Department, California State University Fresno, Fresno, CA 93740 USA}

\date{\today}

\begin{abstract}
A simple, closed-form solution to the Yang-Mills field equations is presented which has a non-Abelian firewall - a spherical ``horizon" where the energy density diverges. By the gravity/gauge duality, this non-Abelian firewall implies the existence of a gravitational firewall. Gravitational firewalls have been proposed as a way of resolving the information loss paradox, but at the cost of violating the equivalence principle. 
\end{abstract}

\maketitle 

Since its initial formulation \cite{hawking}, the information loss paradox \footnote{If a black hole evaporates due to Hawking radiation the information of what formed the black hole is lost. This implies a non-unitary evolution of the system, violating one of the foundational principles quantum mechanics.} has provided physicists with unpalatable choices in regards to its resolution. To address questions raised by the information loss paradox some fundamental aspects of either general relativity or quantum mechanics (or both) have to be given up. The most recent and pointed formulation of the information loss puzzle comes from Almheiri, Marolf, Polchinski, and Sully (AMPS) \cite{AMPS}. AMPS showed that the evaporation of a black hole via Hawking radiation was inconsistent with one of the three foundational principles of modern physics : (i) the unitary evolution of all in-falling matter and out-going Hawking radiation demanded by quantum mechanics, (ii) the validity of local quantum fields theory outside the horizon of a black hole, (iii) the validity of the equivalence principle, the conceptual foundation of general relativity.

One way to address the information loss paradox is through the existence of a firewall - a region of large/infinite energy just below the horizon. A gravitational firewall arises as follows: a virtual particle pair in the vicinity of the event horizon of a black hole is separated, with one of the particles escaping to infinity as Hawking radiation, and the other falling below the event horizon. These particle pairs are entangled since they were created jointly out of the vacuum. One way to solve the information loss paradox is for the Hawking radiation to leak information out of the black hole. The Hawking radiation can accomplish this by becoming entangled with the previously emitted Hawking radiation. But due to the ``monogamy of entanglement" ({\it i.e.} the principle that a particle can only be entangled with one other particle at a time) this implies that the out-going Hawking radiation quanta would need to break its entanglement with its in-going partner, and become entangled with the Hawking radiation emitted earlier. This breaking of the entanglement ``bond" between the in-going and out-going pair would release energy which would pile up just below the horizon, forming a surface of large energy {\it i.e.}  a firewall. This firewall would burn up a free-falling observer as they crossed the horizon. Effectively space-time would end just below the horizon rather than at the central singularity. This would violate the equivalence principle which holds that a free-falling observer should notice nothing unusual at the horizon, which is simply  empty space-time (according the classical GR). If, on the other hand, the initial particle pair remained entangled, then information would be lost  and the evolution of the system would be non-unitary, violating one of the fundamental tenets of quantum mechanics. 

In this essay we present an exact, closed-form solution to a non-Abelian gauge field coupled to a massless scalar field. This non-Abelian solution has the same structure as the gravitational firewall, namely a spherical shell where the energy density of the fields diverges/becomes large. This effectively cuts off the interior region of the shell from the exterior region. We argue, via the gauge/gravity duality conjecture, that the existence of this non-Abelian firewall lends support for the existence of the gravitational firewall, and thus the implied violation of the equivalence principle.  

The non-Abelian system we consider is an $SU(2)$ Yang-Mills field, $W_\mu ^a$ coupled to a massless scalar field $\phi ^a$. Latin indices are $SU(2)$ group indices and Greek indices are space-time indices. The Lagrange density for our system is ${\cal L} = -\frac{1}{4} F^{\mu \nu a} F_{\mu \nu} ^a + 
\frac{1}{2} D^{\mu}  \phi ^a D_\mu \phi^a $  where $F_{\mu \nu} ^a = \partial _{\mu} W_{\nu} ^a - \partial _{\nu} W_{\mu} ^a + g \epsilon ^{abc} W_{\mu} ^b W_{\nu} ^c$ is the field strength tensor. The covariant derivative of the scalar field is  $D_{\mu} \phi ^a = \partial _{\mu} \phi ^a + g \epsilon ^{abc} W_{\mu} ^b \phi ^c$, with $g$ being the $SU(2)$ coupling. 

Our non-Abelian solution is taken to be time-independent and the time component of the gauge field is taken to be zero, $W^a _0 =0$. These conditions imply $\partial _t (\cdots) = 0$ and $D_t (\cdots) = 0$. With these conditions the equations for the gauge and scalar fields become
\begin{equation}
\label{eqn1}
D _i F^{j i a} = g \epsilon ^{abc} (D^j \phi ^b) \phi ^c
~~~~,~~~~
D _i ( D^i \phi ^a )  = 0
\end{equation}
Taking the Wu-Yang ansatz \cite{wu-yang} for the gauge and scalar fields 
\begin{equation}
\label{ansatz}
W_i ^a = \epsilon_{aij} \frac{r^j}{g r^2} [ 1 - K(r)] ~~~~,~~~~
\phi ^a = \frac{r^a}{g r^2} H(r) ~,
\end{equation}
and inserting this into \eqref{eqn1}, yields the following coupled non-linear differential equations for $K(r)$ and $H(r)$
\begin{equation}
\label{nlpdd}
r^2 K'' = K (K^2 + H^2 -1) ~~~~;~~~~
r^2 H'' = 2HK^2
\end{equation} 
It is easy to see the equations \eqref{nlpdd} have the following exact, closed form solution 
\begin{eqnarray}
\label{soln}
K(r) = \frac{C r}{1 - C r} ~~~~;~~~~
H(r) = \frac{ \pm 1}{1 - C r}~,
\end{eqnarray}
with $C$ an arbitrary constant. Equation \eqref{ansatz} shows that the gauge and scalar fields have a Coulomb/$\frac{1}{r}$ singularity at $r=0$. The solution in \eqref{soln} also shows that the gauge and scalar fields develop a singularity on the spherical surface $r=\frac{1}{C}$. It is this feature that we call the non-Abelian firewall.

The solution in equation \eqref{soln} has been discovered/re-discovered periodically since 1977 \cite{protogenov} \cite{lunev} \cite{singleton1}. These works emphasized the similarity of the solution in \eqref{soln} to the classical Schwarzschild solution of general relativity. Just as the fields, $W^a _\mu$ and $\phi ^a$, diverge on the spherical surface $r=\frac{1}{C}$, so too some of the metric and Christoffel components of the Schwarzschild metric, for a mass $M$, diverge at the Schwarzschild radius $r_S=\frac{2GM}{c^2}$. However, there is a difference between the divergence of the non-Abelian firewall solution in \eqref{soln} and the divergence of the classical Schwarzschild solution. The divergence of the Schwarzschild metric at the horizon, $r_S=\frac{2GM}{c^2}$, is well known to be a coordinate singularity ({\it i.e.} an artifact of the coordinate system) rather than a physical singularity. In contrast the singularity in the non-Abelian firewall solution of \eqref{soln} is a real singularity. This is seen by looking at the energy density of the solution from \eqref{soln}. First, we write down the general energy-momentum tensor for the Lagrange density ${\cal L} = -\frac{1}{4} F^{\mu \nu a} F_{\mu \nu} ^a +  \frac{1}{2} D^{\mu}  \phi ^a D_\mu \phi^a $ which is
$T ^{\mu \nu} = \frac{2}{\sqrt{ -g} } 
\frac{ \partial ( {\cal L} \sqrt{ -g} )}{ \partial g_{\mu \nu} } = F^{\mu \beta a} F_{\beta} ^{\nu a} + D^{\mu} \phi ^a D^{\nu} \phi ^a
+ g ^{\mu \nu} {\cal L}$. From this the energy density, $T^{00}$, for the solution from \eqref{soln}, in terms of the ansatz functions $K(r), H(r)$, is  \cite{singleton1}
\begin{equation}
\label{e-density}
 T^{00} = \frac{1}{r^2} \left( {K'} ^2 + \frac{(K ^2 - 1) ^2}{2 r^2}  + \frac{H^2 K^2}{r^2} +
\frac{(rH' - H) ^2}{2 r^2} \right) 
\propto \frac{1}{r^4 (1-Cr)^4} ~.
\end{equation}
The energy density of equation \eqref{e-density} leads to a point like divergence at $r=0$ as well as a divergence on the spherical surface $r =\frac{1}{C}$. The singularity at $r=0$ is a real singularity as is the $r=0$ singularity of the Schwarzschild solution. However the singularity on the spherical surface, $r =\frac{1}{C}$, is fundamentally different from the coordinate singularity at $r_S=\frac{2GM}{c^2}$ of the classical Schwarzschild solution: $r =\frac{1}{C}$ is a real singularity whereas $r_S=\frac{2GM}{c^2}$ is a non-physical, coordinate singularity. The real singularity at $r=\frac{1}{C}$ would seem to be a negative feature. Indeed, the solution in \eqref{soln} is less well known than the finite energy BPS solution \cite{bogo} \cite{PS} precisely because true singularities make the non-Abelian firewall solution of less physical interest, especially from the particle physics point of view.

In this essay we want to argue that this ``sin" of the solution in \eqref{soln} ({\it i.e.} the real singularity at $r=\frac{1}{C}$) is actually a ``virtue", by connecting it to the hypothesized gravitational firewall. As mentioned in the introduction, the AMPS approach to the information loss paradox \cite{AMPS} may lead to the possibility that the horizon, far from being benign, becomes a region with a large energy curtain -- a firewall. This black hole firewall scenario contrasts with the classical picture of a Schwarzschild black hole where a free falling observer should not notice anything unusual at the horizon. The existence of the black hole firewall comes at a cost -- the equivalence principle is violated. If one wants to retain the equivalence principle then one must accept some other unappealing violation of either local quantum field theory or unitarity. 

The existence of the non-Abelian firewall solution in equation \eqref{soln} supports the AMPS firewall solution to the information loss paradox, via the gauge/gravity duality conjecture \cite{maldacena}. Broadly, gauge/gravity duality suggests a connection between gauge theories and gravity. The original, and most precise, formulation of the gauge/gravity duality, due to Maldacena, involves a correspondence between anti-de Sitter on the gravity side and conformal field theory on the gauge side. This is  the celebrated AdS/CFT correspondence. Here we are dealing with neither anti-de Sitter space nor a conformal field theory. Nevertheless, the gauge/gravity duality conjecture proposes that the duality between the gauge theory side and the gravity side is broader than AdS and CFT  \cite{AGMO} -- that the proposed duality exists more generally between gravity and gauge theories. Accepting this broader version of the gauge/gravity duality implies that the existence of the explicit non-Abelian firewall of equation \eqref{soln} supports the existence of the gravitational firewall. Both have the common feature of a large/divergent energy on a spherical surface -- either $r=\frac{1}{C}$ or $r_S=\frac{2GM}{c^2}$.

Using the gauge/gravity duality to support the existence of the gravitational, AMPS firewall via the existence of the non-Abelian firewall in \eqref{soln}, implies accepting the violation of the equivalence principle. There are other lines of work which point toward violation of the equivalence principle when general relativity is combined with quantum mechanics. It has been pointed out \cite{braunstein} that the loss of entanglement of particles across the horizon leads to an ``energetic curtain descending around the black holes". Also, by comparing the details of temperature versus local acceleration for Hawking radiation versus Unruh radiation, one finds \cite{singleton2} a violation of the equivalence principle. The violation of the equivalence principle, when one brings quantum mechanics into the picture, is not a surprise. The equivalence principle is a local principle, while quantum mechanics has inherently non-local features. The non-Abelian firewall solution implies, via the gauge/gravity duality, the existence of the gravitational firewall, and its implied approach to solving the information loss paradox which comes at the cost of violation of the equivalence principle. 


\end{document}